# Transient Stability Assessment of Smart Power System using Complex Networks Framework

A. B. M. Nasiruzzaman, *Member, IEEE*, and H. R. Pota, *Member, IEEE*

*Abstract*—In this paper, a new methodology for stability assessment of a smart power system is proposed. The key to this assessment is an index called betweenness index which is based on ideas from complex network theory. The proposed betweenness index is an improvement of previous works since it considers the actual real power flow through the transmission lines along the network. Furthermore, this work initiates a new area for complex system research to assess the stability of the power system.

*Index Terms*--Transient stability, complex network, smart grid, bus admittance matrix.

## I. INTRODUCTION

MANY of the public infrastructures like electric power network are subject to two types of threats: intentional and accidental [1]. Intentional attack can be subdivided into physical or cyber attacks. According to US government accountability office, in 2002, 70 percent of power companies experienced some kind of severe cyber attack to their computing or energy management systems [2]. Physical attacks against key elements of the system are also common. Whether it is going to be a physical or cyber attack, the modern smart grid must resist. The designers of the modern grid should plan for a dedicated, well planned attack prevention strategy. For the modern grid to resist attack it must reduce the vulnerability of the grid to attack by protecting key assets from cyber, physical, or accidental attacks. The complex networks approach to electric power network security would identify key vulnerabilities, assess the likelihood of threats, and determine the consequences of attacks. One of the particular goals of the security program is to identify critical sites and systems.

Complex networks, which had been the main research area of graph theory, have drawn interest of researchers from various disciplines as graph theory began to focus on statistics and analytics [3]. A complex network was proposed as a random network [4]. There are other networks whose behaviour falls in between regular and random, and these are classified as small-world networks [5]–[7]. A power system usually falls in the small–world network category [8].

Complex network theory has been used to model the power system and analyze its several aspects [8]–[13]. The structural vulnerability of the North American power grid was studied after the August 2003 blackout affecting the United States [14]. Similarly, the large scale blackouts and cascading failures motivated analysis of the Italian power grid based on the model for cascading failures [13], [15]. How redistribution of load on nodes due to failure of certain important nodes can cause a cascading failure was also demonstrated [16], [17].

Vulnerability analysis models [6]–[11] were initially proposed for complex abstract networks and were then used in power systems [1], [7]–[15]. However, those physicists' work neglected some concrete engineering features. Therefore, there are good prospects for researchers to further investigate the complex problems by considering power system characteristics and complex network theory together. Electric power networks are different from these abstract networks. Electric networks are governed by Ohm's and Kirchhoff's Laws which are used to form the bus admittance matrix. These special characteristics result in a unique pattern of interaction between nodes in power grids. Therefore, for better explaining complex blackouts of power systems, an improved model which is based on the system bus admittance matrix is proposed, representing the special electrical topological structure [18].

Till now power system research based on complex network theory has been mainly on fault study or vulnerability studies. Since the nodes and edges of the power grid increase almost every time, and the interaction of the components in power system becomes more and more complex, attention must be paid in new research approaches to solve load flow, fault analysis, and stability analysis problems [3]. Since smart grid adds new dimension and complexity in the power system. In this paper we have proposed a method for addressing transient stability issue in smart power system based on topographical information of power grid.

The rest of the paper is organized as follows. Section II describes a model for analyzing power system within the context of complex networks. Section III describes some statistic parameter for complex network. Section IV gives an index for assessing the stability of complex power system network using complex framework. Some concluding remarks are given in Section V.

A. B. M. Nasiruzzaman and H. R. Pota are with School of Engineering and Information Technology (SEIT) of University of New South Wales at the Australian Defence Force Academy (UNSW@ADFA), Northcott Drive, Canberra, ACT 2600, Australia (e-mail: nasiruzzaman@ieee.org, and h.pota@adfa.edu.au).

## II. Power System as a Complex Network

To analyze the power system within the context of complex network theory, the first step is to model the system as a graph [18]. From the perspective of network theory, a graph is an abstract representation of a set of objects, called nodes or vertices, where some pairs of the objects are connected via links or edges. The power system of today is a complex interconnected network which can be subdivided into four major parts of generation, transmission, distribution and loads [19]. To portray the assemblage of various components of power system, engineers use single-line or one-line diagram which provides significant information about the system in a concise form [20]. Power is supplied form the generator nodes to the load nodes via transmission and/or distribution lines. Since for a given operating condition, power flows only in one direction, a directed graph can be easily constructed from the single-line representation of the power system considering various generators, bus bars, substations, or loads of the system as nodes or vertices and transmission lines and transformers as edges or links between various nodes of the system. The principle of mapping is described as follows:

 a) all impedances between any bus and neutral are neglected,

 b) all transmission and/or distribution lines are modeled except for the local lines in the plants and substations,

 c) all transmission lines and transformers are modeled as weighted lines, the weight is equal to the admittance between the buses, and

 d) parallel lines between buses are modeled as an equivalent single line.

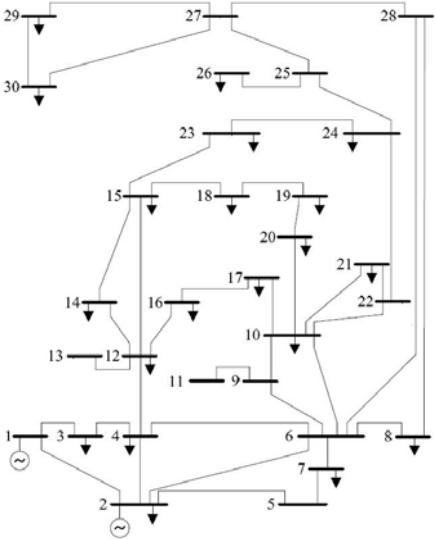

Fig. 1. The IEEE-30 bus system.

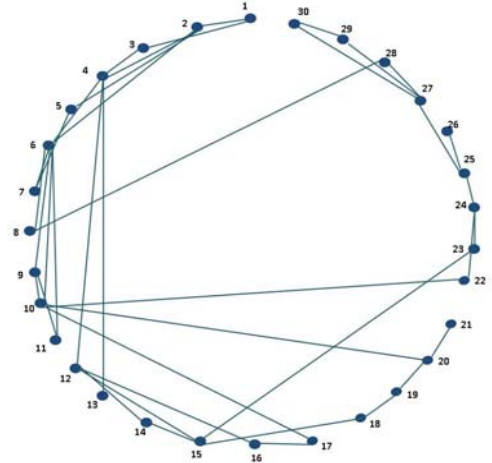

Fig. 2. Physical topology graph of IEEE-30 bus system.

To illustrate mapping of a single-line diagram to a directed graph, a simple example of IEEE 30 bus system [21] is used here. Fig. 1 depicts the IEEE 30 bus system with 30 bus bars, and 41 links connecting them. Fig. 2 is the corresponding mapped graph from the original IEEE 30 bus system. It contains 30 nodes/vertices which correspond to the slack, voltage-controlled, and load bus bars of the original system. The transmission lines are represented by the 41 links/edges which connects various nodes. Now, the weight matrix from the graph has to be formulated. The traditional modeling approach only considers the physical connection [3], [22]–[26], the weight matrix, $W$, (also called adjacency matrix or Boolean matrix) is calculated by considering only the physical topology of the graph. If there is a connection between node $i$ and node $j$ then the corresponding element of the weight matrix $w_{ij} = 1$, otherwise $w_{ij} = 0$. The weight matrix found in traditional model has no sense of directionality, i.e., when nodes $i$ and $j$ are connected $w_{ij} = w_{ji} = 1$. This model does not capture the electrical power system's most important trait like impedance which plays a significant role in the flow of power, losses, stability of the system. Several researchers have considered the reactance of the line [9], [27], neglecting the line resistance which is very small for transmission systems. But, in order to generalize the model for both the transmission and the distribution system, the impedance, (i.e., both the reactance and resistance) needs to be taken into consideration. This approach based on bus admittance matrix is well adopted by various researches in the power system [18], [28]. In this case, the weight matrix can be found from the off-diagonal elements of the bus admittance matrix. For an $n$ bus system the node-voltage equation is written in the matrix form as:

$$\begin{bmatrix} I_1 \\ I_2 \\ . \\ I_i \\ . \\ I_n \end{bmatrix} = \begin{bmatrix} Y_{11} & Y_{12} & . & Y_{1i} & . & Y_{1n} \\ Y_{11} & Y_{12} & . & Y_{1i} & . & Y_{1n} \\ . & . & . & . & . & . \\ Y_{i1} & Y_{i2} & . & Y_{ii} & . & Y_{in} \\ . & . & . & . & . & . \\ Y_{n1} & Y_{n2} & . & Y_{ni} & . & Y_{nn} \end{bmatrix} \begin{bmatrix} V_1 \\ V_2 \\ . \\ V_i \\ . \\ V_n \end{bmatrix} \quad (1)$$

or

$$I_{bus} = Y_{bus}V_{bus} \qquad (2)$$

where, $Y_{bus}$ is the bus admittance matrix. The diagonal elements of the bus admittance matrix correspond to the sum of the impedances of the lines connected to each bus of the system. Since diagonal elements are not included in weight matrix, in effect, the role of various impedances connected from the bus to the neutral is not considered here. The off-diagonal elements are equal to the negative of the equivalent admittance between the nodes. They are known as the mutual or transfer admittances. So, in this case the $ij$–th element of the weight matrix can be found from $w_{ij} = Y_{ij}$. Here, it is obvious that $Y_{bus}$ is a symmetric matrix, i.e., $Y_{ij} = Y_{ji}$. So, the directionality of the power flow is not considered in this model. The information of the direction of power flow within a network can be found from load flow analysis. By conducting power flow we can find the voltage magnitudes and angles of all the buses within the system. If there is a link between bus $i$ and bus $j$, if voltage angle of bus $i$ is higher than that of bus $j$, then power flows from bus $i$ to bus $j$, otherwise power flows in the reverse direction, i.e., from bus $j$ to bus $i$. The weight matrix is constructed using the following rule:

$$w_{ij} = \begin{cases} Y_{ij}, & \text{if } P_{ij} > 0 \\ \infty, & \text{if } P_{ij} \leq 0 \end{cases} \qquad (3)$$

where, $P_{ij}$ indicates the flow of power from node $i$ to node $j$. Fig. 3 shows the directionality of the IEEE 30 bus system in steady-state. Table I summarizes the elements of the weight matrix for IEEE 30 bus system.

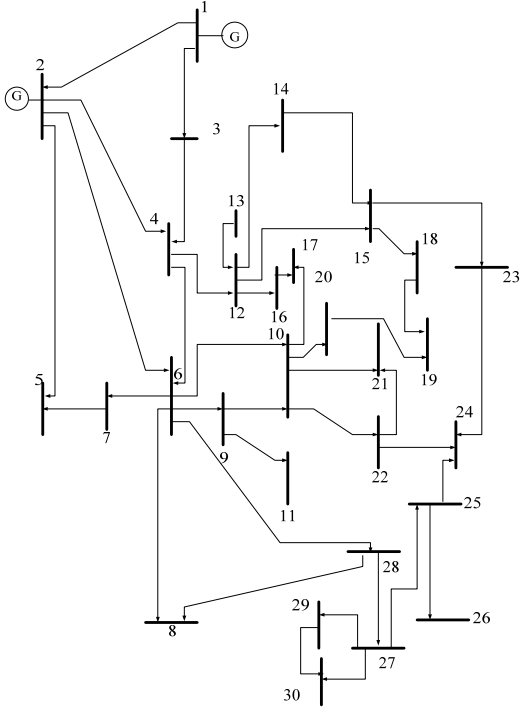

Fig. 3. Power flow diagram of IEEE 30 bus system.

## TABLE I
### Elements of Weight Matrix for IEEE 30 Bus System

| Element | Weight | Element | Weight |
|---------|--------|---------|--------|
| $w_{1-2}$ | 0.0192 + 0.0575i | $w_{12-15}$ | 0.0662 + 0.1304i |
| $w_{1-3}$ | 0.0452 + 0.1852i | $w_{12-16}$ | 0.0945 + 0.1987i |
| $w_{2-4}$ | 0.0570 + 0.1737i | $w_{13-12}$ | 0.0000 + 0.1400i |
| $w_{2-5}$ | 0.0472 + 0.1983i | $w_{14-15}$ | 0.2210 + 0.1997i |
| $w_{2-6}$ | 0.0581 + 0.1763i | $w_{15-18}$ | 0.1073 + 0.2185i |
| $w_{3-4}$ | 0.0132 + 0.0379i | $w_{15-23}$ | 0.1000 + 0.2020i |
| $w_{4-6}$ | 0.0119 + 0.0414i | $w_{16-17}$ | 0.0824 + 0.1923i |
| $w_{4-12}$ | 0.0000 + 0.2560i | $w_{18-19}$ | 0.0639 + 0.1292i |
| $w_{6-7}$ | 0.0267 + 0.0820i | $w_{20-19}$ | 0.0340 + 0.0680i |
| $w_{6-8}$ | 0.0120 + 0.0420i | $w_{22-21}$ | 0.0116 + 0.0236i |
| $w_{6-9}$ | 0.0000 + 0.2080i | $w_{22-24}$ | 0.1150 + 0.1790i |
| $w_{6-10}$ | 0.0000 + 0.5560i | $w_{23-24}$ | 0.1320 + 0.2700i |
| $w_{6-28}$ | 0.0169 + 0.0599i | $w_{25-24}$ | 0.1885 + 0.3292i |
| $w_{7-5}$ | 0.0460 + 0.1160i | $w_{25-26}$ | 0.2544 + 0.3800i |
| $w_{9-11}$ | 0.0000 + 0.2080i | $w_{27-25}$ | 0.1093 + 0.2087i |
| $w_{9-10}$ | 0.0000 + 0.1100i | $w_{27-29}$ | 0.2198 + 0.4153i |
| $w_{10-20}$ | 0.0936 + 0.2090i | $w_{27-30}$ | 0.3202 + 0.6027i |
| $w_{10-17}$ | 0.0324 + 0.0845i | $w_{28-27}$ | 0.0000 + 0.3960i |
| $w_{10-21}$ | 0.0348 + 0.0749i | $w_{27-8}$ | 0.0636 + 0.2000i |
| $w_{10-22}$ | 0.0727 + 0.1499i | $w_{29-30}$ | 0.2399 + 0.4533i |
| $w_{12-14}$ | 0.1231 + 0.2559i | other | $\infty$ |

## III. Topological Statistic Parameter in Power Grid

This section describes some basic statistic parameter of power grid within complex network framework. All of these parameters come from graph theory, the branch of mathematics that deals with networks [29].

### A. Degree

The number of links, directed or undirected, connected with a node $i$ in a graph is called the degree of the node, $d_i$. For the IEEE 30 bus system in Fig. 3 the degree of various nodes are given in Table II. When the graph is directed, the out-degree of a node is equal to the number of outward-directed links, and the in-degree is equal to the number of inward-directed links. The in-degree and out-degree of the IEEE 30 bus system is given in Table III. The hub of a graph is the node with the largest degree. So node 6 with degree 7 is the hub of IEEE 30 bus system. The degree sequence distribution of nodes of the IEEE 30 bus system is shown in Fig. 4.

## TABLE II
### Degree of Various Nodes of IEEE 30 Bus System

| Node | Degree | Node | Degree | Node | Degree |
|------|--------|------|--------|------|--------|
| 1 | 2 | 11 | 1 | 21 | 2 |
| 2 | 4 | 12 | 5 | 22 | 3 |
| 3 | 2 | 13 | 1 | 23 | 2 |
| 4 | 4 | 14 | 2 | 24 | 3 |
| 5 | 2 | 15 | 4 | 25 | 3 |
| 6 | 7 | 16 | 2 | 26 | 1 |
| 7 | 2 | 17 | 2 | 27 | 4 |
| 8 | 2 | 18 | 2 | 28 | 3 |
| 9 | 3 | 19 | 2 | 29 | 2 |
| 10 | 6 | 20 | 2 | 30 | 2 |

| Node | Degree | | Node | Degree | | Node | Degree | |
|---|---|---|---|---|---|---|---|---|
| | In | Out | | In | Out | | In | Out |
| 1 | 0 | 2 | 11 | 1 | 0 | 21 | 2 | 0 |
| 2 | 1 | 3 | 12 | 2 | 3 | 22 | 1 | 2 |
| 3 | 1 | 1 | 13 | 0 | 1 | 23 | 1 | 1 |
| 4 | 2 | 2 | 14 | 1 | 1 | 24 | 3 | 0 |
| 5 | 2 | 0 | 15 | 2 | 2 | 25 | 1 | 2 |
| 6 | 2 | 5 | 16 | 1 | 1 | 26 | 1 | 0 |
| 7 | 1 | 1 | 17 | 2 | 0 | 27 | 1 | 3 |
| 8 | 2 | 0 | 18 | 1 | 1 | 28 | 1 | 2 |
| 9 | 1 | 2 | 19 | 2 | 0 | 29 | 1 | 1 |
| 10 | 2 | 4 | 20 | 1 | 1 | 30 | 2 | 0 |

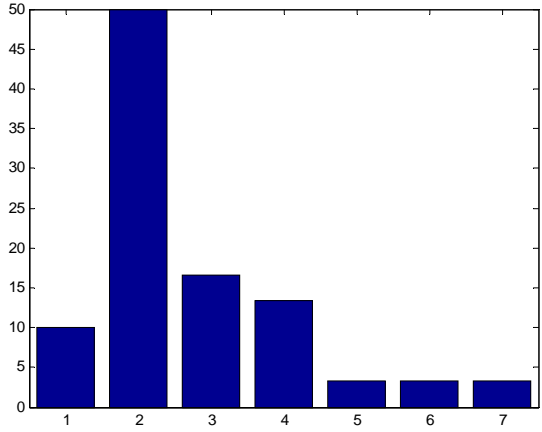

Fig. 4. Degree sequence distribution of IEEE 30 bus system.

### B. Clustering Coefficient

Every node directly connected with a given node is called the neighbor of that node. If there are $d_i$ such neighbors of a node $i$, it means that there may be *[d_i(d_i-1)]/2* potential links among the neighbors of the node $i$. Suppose that the neighbors share $c$ links; then the clustering coefficient of node $i$, *Cc(i)*, is the ratio between actual number of links and maximum possible links.

$$Cc(i) = \frac{2c}{d_i(d_i-1)} \qquad (4)$$

The clustering coefficient of an entire graph is the average over all node clustering coefficients. If there are $n$ nodes in the whole system the clustering coefficient of the whole system or graph $G$, *CC(G)*, is

$$CC(G) = \frac{1}{N}\sum_{i=1}^{N} Cc(i) \qquad (5)$$

The clustering coefficient of IEEE 30 bus system is 0.2348.

### C. Characteristic Path Length

The length of a path is equal to the number of links between starting and ending nodes of the path. Path length is measured in hops—the number of links along the path. The distance between two nodes along a path is equal to the number of hops that separate them. It is possible for a graph to contain multiple paths connecting nodes. Generally, the shortest path is used to calculate the distance between nodes $i$ and $j$. This is also known as the direct path between two nodes. The average path length of a graph is equal to the average over all direct paths. This metric is also known as the characteristic path length of the graph. The diameter of the graph is the maximum distance between any pairs of nodes [3]. The characteristic path length and diameter of the IEEE 30 bus system is 3.43 and 7 hops respectively.

Table IV summarizes various statistic parameters of several IEEE test systems.

| Parameters | 30 Bus | 57 Bus | 118 Bus | 300 Bus |
|---|---|---|---|---|
| Node | 30 | 57 | 118 | 300 |
| Edge | 41 | 80 | 186 | 411 |
| Average Degree of Node | 2.73 | 2.81 | 3.15 | 2.74 |
| Clustering Coefficient | 0.2348 | 0.1211 | 0.1592 | 0.0851 |
| Characteristic Path Length | 3.43 | 5.12 | 2.95 | 5.95 |
| Diameter | 7 | 13 | 9 | 17 |
| Number of Maximum Shortest Path | 3 | 4 | 3 | 1 |

## IV. STABILITY ASSESSMENT OF SMART POWER SYSTEM

A small power system that includes self-contained generation, transmission, distribution, sensors, energy storage, and energy management software is called a micro smart grid [30]. This micro grid has a seamless and synchronized connection to a utility power system but can operate independently as an island from that system. Interconnections are required within several micro smart grids or between today's regional grid layouts and planned renewable energy generators to form future mega grids [30] to transmit electricity to any region where needed. The vision of the grid is to also eliminate congestion problems and balance loads from intermittent energy sources across regions. It is also known as super grid or national grid.

When multiple micro smart grids will be interconnected, they could have a substantial influence on grid stability. Undesirable dynamic interactions could cause key, heavily loaded transmission lines to trip, interrupting power exports and imports between areas. However, if micro grids are designed with their dynamic impact on the transmission system taken into account, i.e., analyzing transient stability beforehand, they can enhance the stability of the transmission lines, which could permit transmission power limits to increase. The transient stability of the system depends on the transfer reactance which is heavily reliant on the topological structure of the power networks. Hence some of the complex network concepts and techniques may be applicable to help analyze the stability of smart grid systems. Ideas from the complex network theory have been used in this paper to find whether a smart power system will be stable or not when subjected to transmission line removal from the system due to fault or overloading.

Research is ongoing on power system vulnerability analysis using complex network theory. There are some critical links in every network which can make the system very vulnerable to attacks. The complex network theory has been used to explain some phenomenon like cascading effects in a power system and identification of vulnerable line. In this paper we address the stability or synchronization issue which is immediate consequence of random or intentional attack on a network by introducing a new vulnerability index called line betweenness which relates to the system stability. Betweenness measures the extent to which a line or edge lies in the shortest paths between various sets of nodes [15]. In order to calculate the betweenness we follow the following steps:

(a) Model the power system as a directed graph from the power flow solution according the mapping procedure described earlier.

(b) Calculate the weight matrix from the mapped directed graph according to (3).

(c) Form a shortest path set including all possible shortest paths from all (source) nodes containing generators to all other nodes using Floyd–Warshall algorithm [31].

(d) Find the betweenness of every line of the directed graph from the shortest path set. If any line is included in the shortest path between generator node $i$ and other node $j$ then the real power flowing in the line is called the betweenness of that line. For the lines that are in multiple shortest paths, add up all the betweenness indices.

(e) Sort and rank the lines according to the betweenness in descending order.

IEEE 30 bus system is analyzed in this manner to find the vulnerability of the system. Table V gives critical lines of the system. To test our hypothesis we performed the multimachine stability analysis of the system. The system is faulted initially and to clear the fault a line is removed from the system at 1 second and the relative swing of the generators with respect to the slack bus is observed to check whether the machines are swinging back to the equilibrium position or going out of sync. It is found that if the lines with high vulnerability are removed the machines cannot maintain synchronism. The lower the vulnerability, the higher is the chance for the post-fault system to be stable.

Table V also compares two different approach of calculating betweenness. In past approach researchers ignored the load of the system [27]. It can be seen from the Fig. 1 of the IEEE 30 bus system that this system consists of only two generators one at bus 1 and other at bus 2. So the impact of removing line 1–3 should be higher than removing lines 6–7, or 6–8. The past approach gives priority to lines 6–7 or 6–8 than the line 1–3 in terms of betweenness index. This is clearly a shortcoming of the past approach since removing line 1–3 would leave only one path to flow power from source to the rest of the system via bus 1 making the system more susceptible to collapse. The proposed approach improves the betweenness of line 1–3 and gives it priority that line 6–7.

To verify our assumption, we simulated the swing equations for this multimachine system and the results are depicted in Figs. 5–7. The simulation results are also tabulated in the third column of the present and past approach of Table V. Transient stability analysis of the network was performed [32]. We can remove any line of the interconnections and see the effect on the relative swings of the machines. The swing equation is the very basic form that we used as given in (6), (7)

$$\frac{d\delta}{dt} = \Delta\omega \qquad (6)$$

$$\frac{d\Delta\omega}{dt} = \frac{\pi f_0}{H}(P_m - P_{max}\sin\delta) \qquad (7)$$

TABLE V
COMPARISON OF BETWEENNESS INDEX

| Proposed Approach | | | Past Approach | | |
|---|---|---|---|---|---|
| Line | Normalized Betweenness | Stability | Line | Normalized Betweenness | Stability |
| $L_{1-2}$ | 1.0000 | Unstable | $L_{1-2}$ | 1.0000 | Unstable |
| $L_{1-3}$ | 1.0000 | Unstable | $L_{2-4}$ | 1.0000 | Unstable |
| $L_{2-4}$ | 1.0000 | Unstable | $L_{2-5}$ | 1.0000 | Unstable |
| $L_{2-5}$ | 1.0000 | Unstable | $L_{2-6}$ | 1.0000 | Unstable |
| $L_{2-6}$ | 1.0000 | Unstable | $L_{6-7}$ | 1.0000 | Stable |
| $L_{6-7}$ | 0.9621 | Stable | $L_{6-8}$ | 1.0000 | Stable |
| $L_{6-8}$ | 0.9621 | Stable | $L_{6-9}$ | 1.0000 | Stable |
| $L_{6-9}$ | 0.9621 | Stable | $L_{6-28}$ | 1.0000 | Stable |
| $L_{6-28}$ | 0.9621 | Stable | $L_{1-3}$ | 0.9635 | Unstable |
| $L_{9-10}$ | 0.4810 | Stable | $L_{9-10}$ | 0.5000 | Stable |
| $L_{9-11}$ | 0.4810 | Stable | $L_{9-11}$ | 0.5000 | Stable |
| $L_{3-4}$ | 0.4000 | Stable | $L_{10-17}$ | 0.3889 | Stable |
| $L_{10-17}$ | 0.3741 | Stable | $L_{10-20}$ | 0.3889 | Stable |
| $L_{10-20}$ | 0.3741 | Stable | $L_{10-21}$ | 0.3889 | Stable |
| $L_{10-21}$ | 0.3741 | Stable | $L_{10-22}$ | 0.3889 | Stable |
| $L_{10-22}$ | 0.3741 | Stable | $L_{3-4}$ | 0.3854 | Stable |
| $L_{4-12}$ | 0.3500 | Stable | $L_{4-12}$ | 0.3372 | Stable |
| $L_{12-14}$ | 0.3207 | Stable | $L_{12-14}$ | 0.3333 | Stable |
| $L_{12-15}$ | 0.3207 | Stable | $L_{12-15}$ | 0.3333 | Stable |
| $L_{12-16}$ | 0.3207 | Stable | $L_{12-16}$ | 0.3333 | Stable |
| $L_{28-27}$ | 0.3207 | Stable | $L_{28-27}$ | 0.3333 | Stable |
| $L_{27-25}$ | 0.2672 | Stable | $L_{27-25}$ | 0.2778 | Stable |
| $L_{27-29}$ | 0.2672 | Stable | $L_{27-29}$ | 0.2778 | Stable |
| $L_{27-30}$ | 0.2672 | Stable | $L_{27-30}$ | 0.2778 | Stable |
| $L_{15-18}$ | 0.1603 | Stable | $L_{15-18}$ | 0.1667 | Stable |
| $L_{15-23}$ | 0.1603 | Stable | $L_{15-23}$ | 0.1667 | Stable |
| $L_{20-19}$ | 0.1069 | Stable | $L_{20-19}$ | 0.1111 | Stable |
| $L_{22-24}$ | 0.1069 | Stable | $L_{22-24}$ | 0.1111 | Stable |
| $L_{25-26}$ | 0.1069 | Stable | $L_{25-26}$ | 0.1111 | Stable |

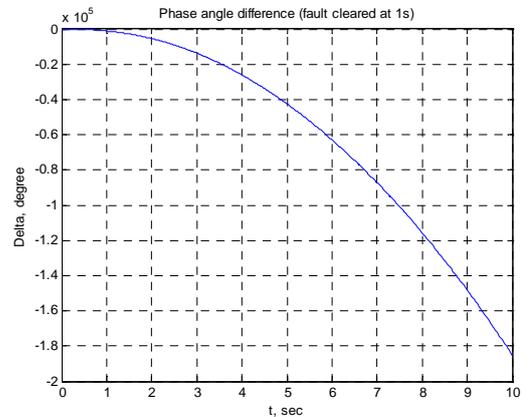

Fig. 5. Transient stability analysis of the IEEE 30 bus system with fault in line 1–2 cleared at 1 sec. Unstable.

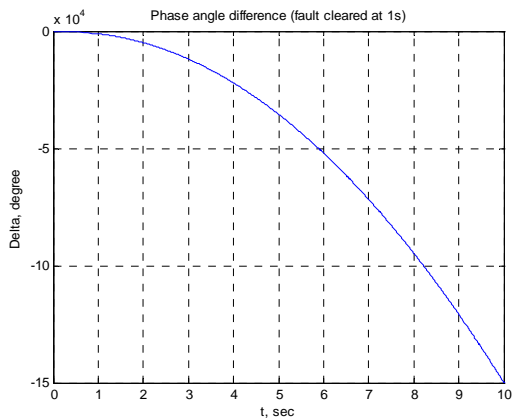

Fig. 6. Transient stability analysis of the IEEE 30 bus system with fault in line 1–3 cleared at 1 sec. Unstable.

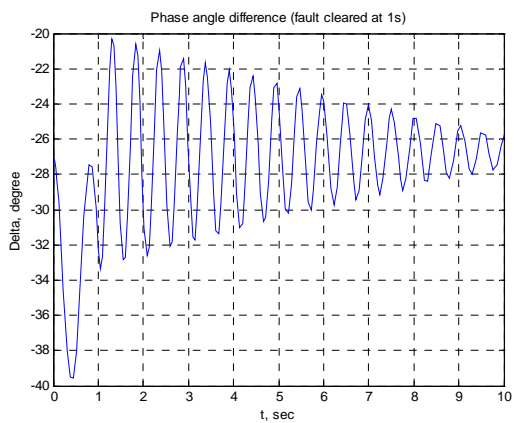

Fig. 7. Transient stability analysis of the IEEE 30 bus system with fault in line 6–7 cleared at 1 sec. Stable.

Next, in order to find the sensitivity of the proposed betweenness index with topology the generator of bus 1 of IEEE 30 bus system is shifted to other buses. This causes change in network topology since changing the generator bus causes a redistribution of power flow. Hence, critical lines of the system change as well. Table VI lists top ten critical lines of the IEEE 30 bus system with generation of bus 1 shifted to buses 3, and 23 respectively.

TABLE VI
SENSITIVITY OF BETWEENNESS INDEX FOR IEEE 30 BUS SYSTEM

| Generator bus 3 | | | Generator bus 23 | | |
|---|---|---|---|---|---|
| Line | Normalized Betweenness | Stability | Line | Normalized Betweenness | Stability |
| L$_{3-1}$ | 1.0000 | Unstable | L$_{23-15}$ | 1.0000 | Unstable |
| L$_{1-4}$ | 1.0000 | Unstable | L$_{23-24}$ | 1.0000 | Unstable |
| L$_{4-2}$ | 1.0000 | Unstable | L$_{15-12}$ | 1.0000 | Unstable |
| L$_{4-6}$ | 1.0000 | Unstable | L$_{15-14}$ | 1.0000 | Unstable |
| L$_{4-12}$ | 1.0000 | Unstable | L$_{15-18}$ | 1.0000 | Unstable |
| L$_{2-6}$ | 0.6786 | Stable | L$_{12-4}$ | 0.7059 | Stable |
| L$_{6-7}$ | 0.6786 | Stable | L$_{12-13}$ | 0.7059 | Stable |
| L$_{6-8}$ | 0.6786 | Stable | L$_{12-16}$ | 0.7059 | Stable |
| L$_{6-9}$ | 0.6786 | Stable | L$_{24-22}$ | 0.7059 | Stable |
| L$_{6-28}$ | 0.6786 | Stable | L$_{24-25}$ | 0.7059 | Stable |

The simulation is repeated for IEEE 57, 118, and 300 bus test system. Tables VII summarizes top fifteen critical lines in IEEE 57, 118, and 300 bus systems. From simulation it can be concluded that there exists a margin in the proposed normalized betweenness index. It can be said that lines with normalized betweenness higher than 0.5 is most critical and can cause stability problem when subject to fault. So, special attention must be given to these critical lines. Again, a line whose normalized betweenness index falls below 0.5 requires less care and maintenance. The normalized betweenness of various IEEE 57 bus test systems is shown in Fig. 8 which also a margin of stability.

TABLE VII
TOP TEN CRITICAL LINES OF VARIOUS STANDARD TEST SYSTEMS

| IEEE 57 Bus System | | IEEE 118 Bus System | | IEEE 300 Bus System | |
|---|---|---|---|---|---|
| Line | Normalized Betweenness | Line | Normalized Betweenness | Line | Normalized Betweenness |
| L$_{1-2}$ | 1.0000 | L$_{9-8}$ | 1.0000 | L$_{2-3}$ | 1.0000 |
| L$_{1-15}$ | 1.0000 | L$_{10-9}$ | 1.0000 | L$_{3-1}$ | 1.0000 |
| L$_{1-16}$ | 1.0000 | L$_{8-5}$ | 0.9697 | L$_{3-4}$ | 1.0000 |
| L$_{1-17}$ | 1.0000 | L$_{8-30}$ | 0.9697 | L$_{3-7}$ | 1.0000 |
| L$_{3-15}$ | 1.0000 | L$_{89-85}$ | 0.8175 | L$_{3-129}$ | 1.0000 |
| L$_{15-13}$ | 1.0000 | L$_{89-88}$ | 0.8175 | L$_{249-3}$ | 1.0000 |
| L$_{15-14}$ | 1.0000 | L$_{89-90}$ | 0.8175 | L$_{4-16}$ | 0.9877 |
| L$_{15-45}$ | 1.0000 | L$_{89-92}$ | 0.8175 | L$_{16-15}$ | 0.9768 |
| L$_{14-46}$ | 0.5820 | L$_{92-91}$ | 0.8175 | L$_{16-36}$ | 0.9768 |
| L$_{46-47}$ | 0.5542 | L$_{92-93}$ | 0.8175 | L$_{33-36}$ | 0.9441 |
| L$_{47-48}$ | 0.5265 | L$_{92-94}$ | 0.8175 | L$_{36-28}$ | 0.9441 |
| L$_{48-38}$ | 0.4988 | L$_{92-102}$ | 0.8175 | L$_{36-35}$ | 0.9441 |
| L$_{38-22}$ | 0.4711 | L$_{49-42}$ | 0.7676 | L$_{36-40}$ | 0.9441 |
| L$_{38-37}$ | 0.4711 | L$_{49-45}$ | 0.7676 | L$_{15-31}$ | 0.6840 |
| L$_{9-13}$ | 0.3325 | L$_{49-47}$ | 0.7676 | L$_{31-32}$ | 0.6840 |

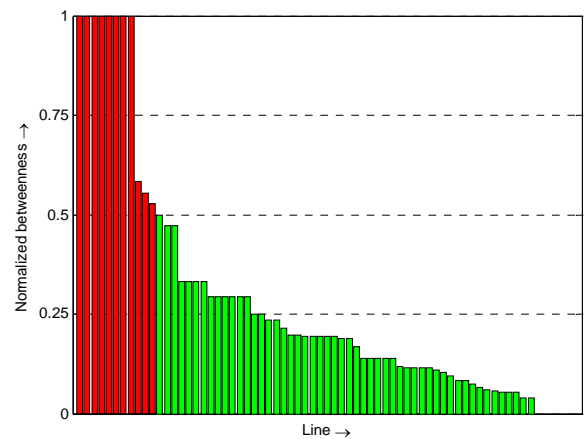

Fig. 8. Normalized betweenness for IEEE 57 bus system.

## V. CONCLUSION

In this paper, we demonstrated the use of complex network theory for vulnerability analysis of power systems after ting into considerations actual electrical parameters. The simulation of various IEEE standard test systems using the method proposed in the paper and other methods used by earlier researchers show clearly that the proposed method is more realistic and draws a margin between stable and unstable region. Although, the proposed approach is simple, it provides a new direction for complex system network research.

## VII. Biographies

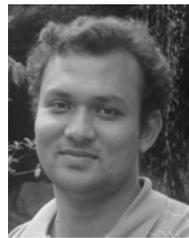

**A. B. M. Nasiruzzaman** (M'2009) was born in Rajshahi in Bangladesh, on December 15, 1983. He received his B.Sc. and M. Sc. Degree in Electrical and Electronic Engineering (EEE) from the Rajshahi University of Engineering & Technology (RUET), Rajshahi, Bangladesh in 2005 and 2008 successively.

He is an assistant professor of EEE in RUET, currently on leave for higher studies at University of New South Wales at the Australian Defence Force Academy (UNSW@ADFA), Canberra, Australia. He is an active member of IEEE and established IEEE RUET student branch in Bangladesh. His research interest is complex network and power system. He developed a MATLAB based power system analysis toolbox which is currently used in RUET power laboratory.

Mr. Nasiruzzaman is also a member of Institution of Engineers Bangladesh and student member of Engineers Australia.

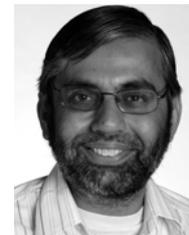

**H. R. Pota** (M'2010) received the B.E. degree from SVRCET, Surat, India, in 1979, the M.E. degree from the IISc, Bangalore, India, in 1981, and the Ph.D. degree from the University of Newcastle, NSW, Australia, in 1985, all in electrical engineering.

He is currently an Associate Professor at the University of New South Wales, Australian Defence Force Academy, Canberra, Australia. He has held visiting appointments at the University of Delaware; Iowa State University; Kansas State University; Old Dominion University; the University of California, San Diego; and the Centre for AI and Robotics, Bangalore. He has a continuing interest in the area of power system dynamics and control, flexible structures, and UAVs.